\documentclass[twocolumn,showpacs,preprintnumbers,amsmath,amssymb]{revtex4}
\usepackage{graphicx}
\usepackage{dcolumn}
\usepackage{bm}
\begin{document}
\title{Effective spin-glass Hamiltonian  
for  the anomalous dynamics of the HMF model}
\author{Alessandro Pluchino, Vito Latora and Andrea Rapisarda}
\affiliation{Dipartimento di Fisica e Astronomia, 
Universit\`a di Catania  and INFN sezione di Catania, 
Via S. Sofia 64, 95123 Catania, Italy\\	}
\date{\today}

\begin{abstract}

We discuss an effective  spin-glass  Hamiltonian 
which can be used to study 
the glassy-like dynamics observed in the metastable states of the Hamiltonian Mean Field (HMF) model.
By means of  the Replica formalism,  we were able to find a self-consistent equation for the 
glassy order parameter which reproduces, 
 in a restricted energy region below the phase transition, the microcanonical simulations for the polarization order parameter recently introducted in the HMF model. 

\end{abstract}
\maketitle
\vspace{0.25cm}


Understanding glassy dynamics 
is one of the greatest challenges in theoretical physics.
Many of the recent  developments 
in this field are based on the analysis of mean-field models\cite{Cugliand, review-glasses}.
The latter  are defined by Hamiltonians with long-range interactions and seem to capture many properties of real systems. 
The rather accurate comparison to numerical simulations~\cite{numerics,Picco} 
and experiments~\cite{experiments,Gris,Ciliberto} 
supports the claim that the  mechanism in these models is similar 
to the one responsible for the glass transition and the glassy dynamics
in real materials.
\\
In this paper we will consider the so-called 
Hamiltonian Mean Field (HMF) model, a system of planar rotators
originally introduced
in Ref.\cite{hmf0}. This model has been intensively
studied in the last years for its extreme richness and
flexibility in exploring the connections between dynamics
and thermodynamics in long-range many-body systems.
In fact, on one hand the model has an exact equilibrium solution,
on the other hand, because of the presence of a kinetic energy
term in the Hamiltonian, the dynamics  can be studied 
by means of molecular dynamics  simulations \cite{hmf0,hmf1,hmf2,lh0}.
From these investigations, many new interesting 
features have emerged which are common to other systems 
with long-range interactions \cite{alfaxy,lj,nobre}.
One of the most intriguing characteristics of the 
dynamics is the existence of quasi-stationary states (QSS), 
i.e. metastable dynamically-created states, whose lifetime 
diverges with the system size $N$ \cite{lrt_pre}. 
In such states, that spontaneously appear in the numerical simulations 
when the system starts from strong off-equilibrium initial conditions, 
many anomalies have been observed, such as 
anomalous diffusion \cite{hmf2},
non-Gaussian velocity distributions \cite{lrt_pre},
vanishing Lyapunov exponents \cite{lrt_pre},
weak ergodicity breaking, hierarchical structures \cite{kyoto}, slow-decaying correlations and aging \cite{mat,phd,yama-ruffo,celia-tama}.
These features have suggested a possible  application of
Tsallis generalized thermodynamics \cite{Tsa,lh1,lrt_pre,moya,cho} 
but also an interesting link with glassy dynamics.
\\
We have shown in previous papers that, in the QSS regime, the HMF system behaves very similary to a spin-glass system \cite{pre,pha,kyoto}. 
Actually, by means of a new order parameter called {\em polarization} and inspired  by the  Edwards-Anderson spin-glass order parameter \cite{EA1,EA2,SK1,SK2}, it has been possible to characterize the dynamically generated QSS as a sort of glassy phase of the HMF model, despite the fact that
neither quenched disorder nor frustration are 
present in the interactions. 
\\
In this paper, by means of a Replica-Symmetry analysis performed on an appropriate effective Hamiltonian, we will show that it is possible to find out a self-consistent equation for a spin-glass order parameter describing, in the thermodynamic limit, the quenched dynamics observed in the QSS regime. We will also show  that the solutions of this equation  reproduce well the microcanonical simulations results for the polarization in the energy region where the dynamical anomalies are more evident, thus strongly suggesting the identification of the two order parameters and confirming the interpretation of the limiting QSS regime as a glassy phase.

\section{Spin-glass Models}

In the last three decades 
spin-glasses have attracted the 
attention of experimentalists and theoreticians as glassy prototypical 
systems showing {\it frustration and quenched disorder}~\cite{spin-glasses-old,Levetal,Mepavi,Kawamura,bou,bert}. 
Shortly, spin-glasses are systems with localized electronic magnetic moments whose
interactions are characterized by quenched randomness: a given
pair of spins have a roughly equal {\it
a priori\/} probability of having a ferromagnetic or an antiferromagnetic
interaction.  The prototype material is a dilute magnetic alloy, with a
small amount of magnetic impurity randomly substituted into the lattice of
a nonmagnetic metallic host. 
In this situation, the impossibility
to minimize simultaneously the interaction energies of all the
couple of spins leads to a frustration which determines
a very complex energetic landscape in phase space. The latter
appears as consisting of large valleys separated by high
activation energies. Each valley contains many local minima in
which the system, at low temperature, can remain trapped for a
very long time. This time grows exponentially with the height of
the energy barriers, thus the system shows very slow relaxation,
strong memory effects and aging.
\\
The modern theory of spin glasses \cite{Stein} began in 1975 with the work of Edwards and
Anderson (EA) \cite{EA1}, who proposed that the essential physics of spin
glasses lays not in the details of their microscopic interactions but rather
in the {\it competition} between quenched ferromagnetic and
antiferromagnetic interactions (i.e. in the frustration). 
Thus they proposed a short range simplified model for spin-glasses, in which one represents the magnetic impurities with Ising spins $s_i=\pm 1$ placed on the vertices of a three dimensional cubic lattice. 
The random nature of the interactions are mimicked with
first neighbors random interactions between the spins 
taken from a Gaussian probability distribution with 
zero mean and variance $J_{ij}^2={\tilde J}^2/(2z)$ 
where $z$ is the connectivity  of the lattice. 
The Hamiltonian (in the absence of an external magnetic field) is 
\begin{equation}
H_J[{\vec S}] = - \sum_{\langle ij\rangle} J_{ij} s_i s_j \; ,
\label{3DEA-hamil}
\end{equation}
where the vector $\vec S$ encodes the full set of spins in the 
sample $\vec S=(s_1,s_2,\dots,s_N)$ and 
${\langle ij\rangle}$ represents nearest neighbors on the lattice.
Shortly after the  appearance of the EA model, an infinite-ranged version was
proposed by Sherrington and Kirkpatrick (SK) \cite{SK1,SK2}.  For a system of
$N$ Ising spins, and in zero external field, the SK Hamiltonian is
\begin{equation}
H_J[{\vec S}] = - {1\over\sqrt{N}} \sum_{(i,j)} J_{ij} s_i s_j  
\; ,
\label{sk-hamil}
\end{equation}
with a Gaussian distribution of interactions.
Notice that in Eq.\ref{sk-hamil} the sum runs over all pair of spins and the factor
$1/\sqrt{N}$ allows to consider the thermodynamic limit for the free
energy and for other thermodynamic quantities.
In ref.\cite{SK2} SK showed that their model has an equilibrium phase transition to a spin-glass phase below the
temperature $T_c=1$ and for an opportune choice of the parameters
$J_0$ and $J$, respectively mean and standard deviation of the Gaussian distribution of interactions.
\\
A mean field theory, employing the Onsager reaction field term, was
proposed two years later by Thouless, Anderson, and Palmer \cite{TAP77},
which indicated that there might be many low-temperature solutions corresponding
to different spin-glass phases.
But the correct solution for the
low-temperature phase of the SK model, due to Parisi
\cite{P79}, employed a novel {\it ansatz\/} and required several more years
before a physical interpretation could be worked out
\cite{P83}. The picture that finally arose was that of a
system with an extraordinary new kind of symmetry breaking, known today as
"Replica Symmetry Breaking", or RSB, after the mathematical procedures used
to derive it.  The essential idea is that the low-temperature phase
consists not of a single spin-reversed pair of states, but rather of
"infinitely many pure thermodynamic states" \cite{P83}, not related by
any simple symmetry transformation. 
\\
Within the original mean-field framework of the SK model
it is possible to observe three different phases, namely, paramagnetic (PA), ferromagnetic (FE) and spin glass (SG) phase, depending on the temperature and the parameters of the Gaussian distribution of the interactions \cite{EA2}.
Each phase is
characterized by a different microscopic behavior and a different
kind of orientation order, giving rise to a characteristic value of the usual mean field order parameter, i.e. the magnetization $m(T)$.
Thus it is clear that the magnetization, calculated at one instant of time, vanishes in the SG phase just like in the PA one.
Therefore, in order to discriminate between spin glass disorder
and paramagnetism, one needs an additional order parameter.
Such a parameter was originally proposed in refs. \cite{EA1,EA2}. 
It is called {\it 'Edwards-Anderson (EA) order parameter'} and 
takes into account the temporal evolution of each
spin. In this way the latter is  able to measure the degree of freezing of the system and to distinguish between the PA phase, where it vanishes together with the magnetization, and the SG phase, where it remains finite.
Actually, in its original formulation, the EA order parameter results to be only an approximation (the so-called {\it Replica Symmetry Approximation}) of the true spin-glass order parameter proposed by Parisi \cite{P79}, that has to be defined in the 'replica space' and results to be a whole function  (see also refs.\cite{bou,Mepavi} for more details).
However, we will show that also considering the approximated SK point of view it is possible
to shed new light on the already suggested link between glassy dynamics and the anomalous out-of-equilibrium behavior of the Quasi Stationary States of the HMF model.

\section{Dynamical Frustration and Polarization in the HMF model}

The HMF model describes a system of $N$ fully-coupled classical inertial
XY spins with unitary module \cite{hmf0}
\begin{equation}
\label{spin}
\stackrel{\vector(1,0){8}}{s_i} = (cos~\theta_i,sin~\theta_i)~~~~~~i=1,...,N~~.
\end{equation}
For a better visualization these spins can also be imagined as particles rotating 
on a unit circle. 
The equations of motion follow from the  Hamiltonian
\begin{equation}
\label{hamiltonian}
        H
= \sum_{i=1}^N  {{p_i}^2 \over 2} +
  { 1\over{2N}} \sum_{i,j=1}^N  [1-cos( \theta_i -\theta_j)]~~,
\end{equation}
where ${\theta_i}$ ($0 < \theta_i \le 2 \pi$) is the angle 
and $p_i$ the conjugate variable representing the rotational 
velocity of spin $i$.  
\\ 
The equilibrium solution of the model in the canonical ensemble 
predicts a second order phase transition from a high 
temperature paramagnetic (PA) phase to a low temperature 
ferromagnetic (FE) one \cite{hmf0,hmf1,hmf2,lh0}. 
The critical temperature is $T_c=0.5$ and corresponds to
a critical energy per particle $U_c = E_c /N =0.75$.
The order parameter of this phase transition is the modulus of
the {\it average magnetization} per spin defined as:
$M = {1\over{N}} | \sum_{i=1}^N
\stackrel{\vector(1,0){8}}{s_i} | ~~$.
Above $T_c$, in the PA phase, the spins point in different 
directions and $M \sim 0$. 
Below $T_c$, in the FE phase, all the spins
are aligned (the rotators are trapped in a single cluster) 
and $M \neq0$.  
\begin{center}
\begin{figure}
\includegraphics[width=3.2in,angle=0]{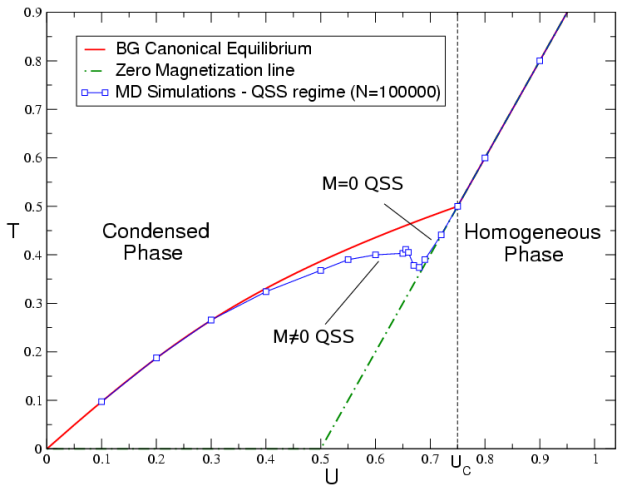}
\caption{\label{fig1} 
In this figure we plot - for N=100000 - the molecular dynamics numerical results corresponding to QSS  (open squares), compared with both the canonical caloric curve, plotted as solid line, and the zero magnetization  (or minimum temperature)  line, reported as dot-dashed. The temperature is calculated by means of the average kinetic energy K,i.e.  $T=2K/N$.  The QSS follows the latter line only up to $U=0.68$, becoming unstable below this limiting value \protect\cite{yama-ruffo,celia-val}.}
\end{figure}
\end{center}

However, as already pointed out in the introduction, molecular dynamics simulations, at fixed energy and for out-of-equilibrium initial conditions, show the presence of long-living quasi-stationary states (QSS) in the subcritical energy density region $0.5\le U \le U_c$ (\cite{hmf1,lrt_pre,phd}). In the \textit{thermodynamic limit } - when they become stationary and reach a limiting temperature value $T_{QSS}(U)$ - the QSS can be usefully divided in two subset of states, depending on their magnetization:
\\
- \textit{$M=0$ QSS} (for $0.68\le U \le U_c$), i.e. anomalous states lieing, for $N \rightarrow \infty$, over the extension of the high temperature branch of the caloric curve (see Fig.\ref{fig1}) and showing a vanishing magnetization ($M=0$). It should be stressed that, because of the energy density-temperature relationship (\cite{hmf0}) :
\begin{equation}
\label{caloric}
U = \frac{T}{2}+\frac{1}{2} (1 - M^2) ~~,
\end{equation}
the zero magnetization constraint forces the system to follow the minimum kinetic energy (maximum potential energy) line, i.e. the minimum temperature line (dot-dashed in Fig.\ref{fig1}); 
\\
- \textit{$M\neq0$ QSS} (for $0.5\le U < 0.68$), i.e. anomalous states with a macroscopic magnetization ($M\neq0$), which for $N \rightarrow \infty$ tend to rejoin the canonical equilibrium curve. 
\\
It has also been shown \cite{yama-ruffo,celia-val} that the {\it M=0 QSS} family is unstable below the limiting energy density value $U \sim 0.68$, corresponding to a temperature $T_{QSS}=0.36$, around which the anomalies are more evident (this is also the reason why, historically, the QSS analysis focused on the value $U=0.69$). Below such a value the QSS cannot follow the zero magnetization (minimum temperature) line and the dynamical anomalies start to decrease until they completely disappear below $U=0.5$.
\\
Actually, it was just the vanishing magnetization of the {\it M=0 QSS} and the discovery of aging ~\cite{mat} and of \textit{dynamical frustration} ~\cite{phd}, i.e. the formation of clusters of particles with power-law size distributions \cite{kyoto} that suggested the interpretation of such a regime in terms of a sort of spin-glass phase characterized by an EA-like order parameter \cite{pre}. 
\\
During the QSS regime, the mean-field interaction seems to be broken: in fact, the clusters that compete in trapping the particles on the unitary circle, generate a dynamically frustrated scenario, in wich each particle never feels all the other particles, but only a restricted number of them. This feature seems to indicate, for finite sizes of the system, a corrugated potential landscape for the single particle. This effect is very sensitive to the initial conditions \cite{kyoto}.
In the thermodynamic limit (when the QSS become stationary), the force between the particles - which depends on M - vanishes and the dynamics is quenched (apart from the global motion imposed by the conservation of total momentum).
\\
Therefore, inspired by the physical meaning of the EA order parameter,
we proposed \cite{pre} a new order parameter for the HMF
model, the so-called {\it polarization}, in order to measure the degree of freezing of the particles (rotators) in the QSS regime and, thus, to characterize in a quantitative way the emerging glassy-like dynamical frustration.
\\
Polarization is defined as the following spatial average
\begin{equation}
\label{pol}
{\it p}={1\over{N}} \sum_{i=1}^N  | <\stackrel{\vector(1,0){8}}{s_i}>|~~~~,
\end{equation}
where
\begin{equation}
\label{elpol}
<\stackrel{\vector(1,0){8}}{s_i}>={1\over{\tau}} \int_{t_0}^{t_0+\tau}
\stackrel{\vector(1,0){8}}{s_i}(t)dt~~~~~~i=1,...,N ~~
\end{equation}
is the {\it elementary polarization}, defined as the temporal
average, integrated over an opportune time interval $\tau$, of the
successive positions of each rotator.
\\
For the typical energy density value $U=0.69$ (corresponding to a limiting temperature $T_{QSS}=0.38$), we showed that, while the magnetization correctly vanishes in the thermodynamic limit, the polarization remains approximatively constant and different from zero, thus quantifying the freezing of the rotators in the QSS regime. 
On the other hand, as shown in ref. \cite{pre,pha}, for $U>U_c$ the polarization coincides with the magnetization and goes to zero.
\\
In other words, the polarization seems to play here the same role played by the Edwards-Anderson \cite{SK1,SK2} order parameter $q_{EA}$ in the SK model, thus characterizing  the anomalous QSS regime, which becomes stable in the thermodynamic limit, as a sort of \textit{Spin-glass phase} for the HMF model.

\begin{center}
\begin{figure}
\includegraphics[width=3.2in,angle=0]{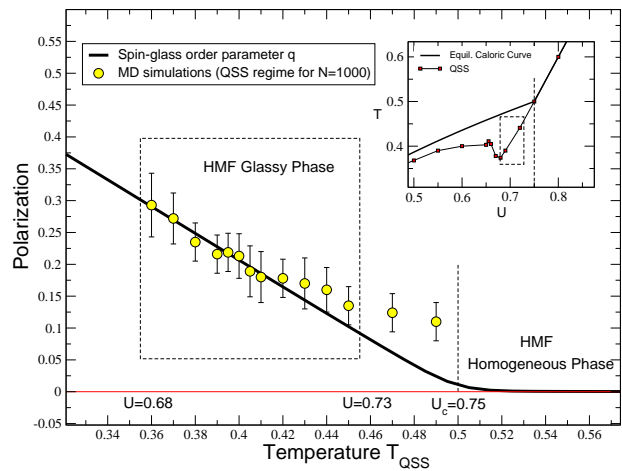}
\caption{\label{fig2} 
Microcanonical simulations for the polarization, Eq.(6), performed in the QSS regime of the HMF model (open circles), are compared with  
 the solution of the self-consistent equation for the spin-glass order parameter $q$ (full line) Eq.(\ref{p_equat}).
Both these quantities are  plotted versus the limiting temperature $T_{QSS}$. 
In the inset, a plot of the caloric curve and the QSS points for the HMF model is reported for comparison. The spin-glass phase region has been framed.}
\end{figure}
\end{center}

In Fig.\ref{fig2}, extending the results reported  in ref. \cite{pre,pha} for $U=0.69$, we  plot, as open circles, the values of the polarization order parameter for  the range $0.68 < U < U_c (U_c=0.75)$, where the {\it M=0 QSS} family result to be stable. These points, reported for convenience as a function of the corresponding $N \rightarrow \infty$ limiting  temperature $T_{QSS}$ ($0.36 < T_{QSS} < T_c=0.5$) were obtained by means of the usual microcanonical molecular dynamics  simulations for $N=1000$. The standard  "water-bag" initial conditions with initial magnetization $M=1$ (all the angles equal and velocities uniformly distributed according to the available kinetic energy) were considered. As usual, the integration time $\tau$ ($\tau=2000$) has been chosen in order to stay inside the QSS plateaux for every $N$ \cite{pre,pha,kyoto}. An average over 25 events was also performed. The error bars represent the fluctuations of the elementary polarization  over the configuration of the $N$ rotators. As previously seen for $U=0.69$ \cite{kyoto}, increasing the size of the system the average polarization remains almost constant inside this error along the QSS plateaux.
Please note that for $0.3<U<U_c$ the system is in the so-called {\it translational regime} and its center of mass drifts \cite{hmf0}, i.e.
 the system is always moving with a global resulting motion
that can be expressed by the phase of the average magnetization. 
Then, as done in the previous calculations \cite{pre,pha},
in order to compute the elementary polarization, one has  to subtract this phase from
the spin angles. 
As expected, the polarization correctly decreases approaching the phase transition, after which, in the homogeneous phase (not shown), it vanishes - together with the magnetization - in the thermodynamic limit. 
\\
In the following sections we give further support to 
the claim that the glassy-like behavior characterizing the QSS regime corresponds to a  2-vector SK Spin-Glass phase by means of a  Replica Method formalism applied to an appropriately chosen Hamiltonian.

\section{The Replica Method}

Suppose \cite{Mepavi} that we have a system characterized by an Hamiltonian $H_J[\overrightarrow{s}]$ depending on the configuration [$\overrightarrow{s}$] of the $N$ spins and on some \textit{quenched variables} J's changing on a time scale infinitely larger than the $\overrightarrow{s}$'s. 
If we also suppose that these control variables are distributed according to a given probability distribution P[J], 
for each choice of the J's one can calculate the partition function
\begin{equation}
Z_J = \sum_{\left\lbrace s \right\rbrace } \exp{(-\beta H_J[\overrightarrow{s}])}~~~~,
\label{zj}
\end{equation}
and the free energy density
\begin{equation}
f_J = -\beta^{-1} \lim_{N \to \infty} \frac{\ln Z_J}{N}~~~~.
\label{free_en}
\end{equation}
The point is that averaging the free energy density over the distribution P[J], i.e.
\begin{equation}
\overline{f_J} = \sum_J P[J]f_J~~.
\label{free_en_av}
\end{equation}
results to be a task not simple at all.
Thus in refs \cite{EA1,EA2} the so called \textit{Replica Method} was  proposed. The latter 
 is a trick to simplify the calculation of Eq.(\ref{free_en_av}) and
consists in computing the average of the free energy density by some analytic continuation procedure from the average of the partition function of \textit{n} uncoupled replicas of the initial system \cite{Mepavi}. 
In fact, using the identity 
\begin{displaymath}
\ln Z_J = \lim_{n \to 0} \frac{(Z_J)^n - 1} {n}~~~~,
\end{displaymath}
togheter with Eq.(\ref{free_en}), Eq.(\ref{free_en_av}) can be rewritten as 
\begin{displaymath}
f = \overline{f_J} = -\beta^{-1} \lim_{N \to \infty} \frac{1}{N} \sum_J P[J] \ln Z_J 
\end{displaymath}
\begin{equation}
= -\beta^{-1} \lim_{N \to \infty} \lim_{n \to 0} \frac{1}{nN} \sum_J P[J] \{(Z_J)^n - 1\}~~~.
\label{ff}
\end{equation}
Finally, if we define
\begin{equation}
\overline{(Z_J)^n} =  \sum_J{P[J] \lbrace Z_J \rbrace^n}
\label{zn}
\end{equation}
and we use the normalization condition $\sum_J P[J]=1$, 
Eq.(\ref{ff}) becomes
\begin{equation}
f = -\beta^{-1} \lim_{N \to \infty} \lim_{n \to 0} \frac{1}{nN} \{\overline{(Z_J)^n} - 1\} 
\label{free_en_new}
\end{equation}
\\
Now, denoting with $a$ the replica index ($a=1,...n,$ with integer $n$), we can write the partition function in Eq.(\ref{zn}) 
as the partition function of $n$ non-interacting replicas of the same system (for the same set of J's)
\begin{displaymath}
(Z_J)^n = \sum_{\lbrace s^1\rbrace}  \sum_{\lbrace s^2 \rbrace}. . . \sum_{\lbrace s^n \rbrace} \exp \lbrace -\sum_{a=1}^n \beta H_J [\overrightarrow{s}^a]\rbrace 
\end{displaymath}
\begin{equation}
= Tr \exp \lbrace -\sum_{a=1}^n \beta H_J [\overrightarrow{s}^a]\rbrace ~~~~,
\label{zn2}
\end{equation}
where the trace $Tr$ in the last expression syntethizes the sums over the spins in all the replicas. 
Further averaging this quantity over the distribution $P[J]$, the calculation of the averaged free energy density follows straightforward from Eq.(\ref{free_en_new}).
\\
Our idea is to describe the glassy dynamics of the {\it M=0 QSS} regime of the HMF model by
studying the equilibrium properties of an infinite range XY spin-glass effective Hamiltonian
\begin{equation}
H_J[\overrightarrow{s}] = -\frac{1}{2}\sum_{i,j}J_{ij}\overrightarrow{s_i}\cdot\overrightarrow{s_j}    ~~~~,
\label{Hj}
\end{equation}
with an opportune choice of the interactions $J_{ij}$. 
In Eq.(\ref{Hj}) each spin (rotator) is defined as $\overrightarrow{s_i}=(cos\theta_i, sin\theta_i)$, with $0<\theta_i<2\pi$ and unitary module, while the factor $\frac{1}{2}$ before the summation prevents from counting two times the same spin couples.
The distribution of the quenched variables $J_{ij}$ has to be chosen such that:
\begin{itemize}
\item
it must take into account the presence of the dynamically created clusters of particles observed in the QSS regime, which in turn generate dynamical frustration; 
\item
its first moment $J_o$ should be equal to $1/N$ (being in the HMF model $J_{ij}=1/N~ \forall{i,j}$); in such a way, for $N \rightarrow \infty$, Eq.(\ref{Hj}) will reduce to the potential term of the HMF model.
\end{itemize}
Thus, without loss of generality, we can choose a Gaussian distribution:
\begin{equation}
p(J_{ij})=[(2\pi)^{1/2}J]^{-1}~exp\frac{-(J_{ij}-J_0)^2}{2J^2}~~
\label{Jdistr}
\end{equation}
where the first two moments, the mean $J_o$ and the variance $J^2$, will be set equal to $1/N$. 
\\
Finally, the inverse temperature $\beta=1/T$ will be fixed by the kinetic term of the HMF model Hamiltonian (being $T=2K/N$), that we assume to be constant for the canonical procedure we want to perform.
Of course we are implicitly assuming that the stationarity of the {\it M=0 QSS} regime in the thermodynamic limit could enable us to use equilibrium thermodynamics tools.
\\
For fixed J's we expect that the replica method would enable us to find out a self-consistent equation for the spin-glass order parameter of the model (\ref{Hj}). 
This equation will be obtained using the Replica-Symmetry (RS) ansatz in the context of a 2-vector infinite range spin-glass model \cite{SK2}, by imposing the spin-glass extremal constraints during the steepest descent procedure.
Our final goal will be to compare such a theoretical prediction with the molecular dynamics results for the polarization of Eq.(\ref{pol}), in a range of temperatures corresponding with the $N\rightarrow\infty$ limiting temperatures ($T_{QSS}$) that characterize the homogeneous quasi-stationary states. 
\\
Let us start by applying the replica trick to the free energy calculation starting from the effective Hamiltonian (\ref{Hj}), with quenched couplings $J_{ij}$ following the distribution (\ref{Jdistr}). 
\\
Using some general properties of the characteristic function of a statistical distribution, and averaging Eq.(\ref{zn2}) over the chosen distribution of the $J_{ij}$, we can write the free energy density expression (\ref{free_en_new}) as a function of only the first two moments of the distribution itself 
\begin{displaymath}
f = -\beta^{-1} \lim_{N \to \infty}\lim_{n \to 0} \frac{1}{nN}
\end{displaymath}
\begin{displaymath}
\times \{ Tr \exp\sum_{(ij)} ~(~\beta J_o \sum_a \overrightarrow{s_i}^a \cdot \overrightarrow{s_j}^a 
\end{displaymath}
\begin{equation}
+ \beta^2 \frac{J^2}{2} \sum_{a}
\overrightarrow{s_i}^a \cdot \overrightarrow{s_j}^a
\sum_b \overrightarrow{s_i}^b \cdot \overrightarrow{s_j}^b ~) - 1 \}
\label{free_en_1}~~,
\end{equation}
where $a,b$ are replica indexes and $i,j$ are spin indexes in each replica. Thus $\overrightarrow{s_i}^a = (cos{\theta_i}^a, sin{\theta_i}^a)$ represents the $i$th spin in the $a$th replica.
As seen before, both the parameters $J_o$ and $J^2$, respectively mean and variance of the J's distribution (\ref{Jdistr}), have to be put equal to $1/N$. 
Finally, the notation $(i,j)$ in the sum is equivalent to a sum over all $N(N-1)/2$ distinct pairs of sites, thus the factor $1/2$ disappears behind the sum over $i,j$.
\\
Because of the latter notation, and after some rearrangment, we can write the following equivalences
\begin{equation}
\sum_{(ij)} \sum_a  \overrightarrow{s_i}^a \cdot \overrightarrow{s_j}^a \frac{1}{2} \sum_a { \vert \sum_i \overrightarrow{s_i}^a \vert ^2} - \frac{nN}{2}
\label{s1}
\end{equation}
and
\begin{displaymath}
\sum_{(ij)} {\sum_{a} \overrightarrow{s_i}^a \cdot \overrightarrow{s_j}^a
\sum_{b} \overrightarrow{s_i}^b \cdot \overrightarrow{s_j}^b }
\end{displaymath}
\begin{displaymath}
= \sum_{a,b} {\sum_{(i,j)} cos(\theta_i^a - \theta_j^a) cos(\theta_i^b - \theta_j^b)} 
\end{displaymath}
\begin{equation}
=\sum_{a,b} {\sum_{(i,j)} \frac{1}{2}
[\overrightarrow{S_i}^{ab} \cdot \overrightarrow{S_j}^{ab} +
\overrightarrow{T_i}^{ab} \cdot \overrightarrow{T_j}^{ab}]}
\label{ss1}
\end{equation}
where two terms of interference between replicas appear
\begin{displaymath}
\overrightarrow{S_i}^{ab}=(cos({\theta_i}^a - {\theta_i}^b), sin({\theta_i}^a - {\theta_i}^b))~~,
\end{displaymath}
and
\begin{displaymath}
\overrightarrow{T_i}^{ab}=(cos({\theta_i}^a + {\theta_i}^b), sin({\theta_i}^a + {\theta_i}^b))~~.
\end{displaymath}
The latter term, for $a=b$, becomes $\overrightarrow{U_i}^{a}=(cos2{\theta_i}^a,sin2{\theta_i}^a)$.
\\
After further rearrangment in the summations, we obtain the following expression for the free energy density
\begin{displaymath}
f = -\beta^{-1} \lim_{N \to \infty}\lim_{n \to 0} \frac{1}{nN}~
\{ \exp~\left[\frac{\beta^2}{8} (n N - 2 n^2) - \frac{n \beta}{2}\right]
\end{displaymath}
\begin{displaymath}
\times Tr \exp~[~\frac{\beta}{2N}\sum_a { \vert \sum_i \overrightarrow{s_i}^a \vert ^2} +
\end{displaymath}
\begin{displaymath}
\frac{\beta^2}{8N} \left( \sum_a { \vert \sum_i \overrightarrow{U_i}^a \vert ^2 + \sum_{a\neq b}({ \vert \sum_i \overrightarrow{S_i}^{ab} \vert ^2} + { \vert \sum_i \overrightarrow{T_i}^{ab} \vert ^2})} \right) ~] -1 \}~~.
\end{displaymath}
\begin{equation}
\label{free_en_2}
\end{equation}
In the thermodynamic limit, the first exponential becomes
\begin{displaymath}
\exp~[\frac{\beta^2}{8} (n N - 2 n^2) - \frac{n \beta}{2}] \approx
\exp~[\frac{nN \beta^2}{8}]
\end{displaymath}
and it does not involve the glassy properties of the system.
Therefore in the following we will concentrate on the term
\begin{displaymath}
I = Tr \exp~[~\frac{\beta}{2N}\sum_a { \vert \sum_i \overrightarrow{s_i}^a \vert ^2} +
\end{displaymath}
\begin{displaymath}
\frac{\beta^2}{8N} \left( \sum_a { \vert \sum_i \overrightarrow{U_i}^a \vert ^2 + \sum_{a\neq b}({ \vert \sum_i \overrightarrow{S_i}^{ab} \vert ^2} + { \vert \sum_i \overrightarrow{T_i}^{ab} \vert ^2})} \right) ~] ~~.
\end{displaymath}
\begin{equation}
\label{I}
\end{equation}
It can be linearized, term by term, with the Hubbard-Stratonovich (HS) Gaussian transformation, i.e.
\begin{equation}
\exp{\left[\mu x^2\right]}(2\pi)^{-1/2} \int
dy \exp{[- \frac{{y}^2}{2} + (2 \mu)^{1/2}xy~]}~~,
\label{HS1}
\end{equation}
with the positions
\begin{displaymath}
\overrightarrow{S}^a = \frac{1}{N} \sum_i \overrightarrow{s_i}^a~~,
\overrightarrow{U}^a = \frac{1}{N} \sum_i \overrightarrow{U_i}^a~~,
\end{displaymath}
\begin{displaymath}
\overrightarrow{T}^{ab} = \frac{1}{N} \sum_i \overrightarrow{T_i}^{ab}~~,
\overrightarrow{S}^{ab} = \frac{1}{N} \sum_i \overrightarrow{S_i}^{ab}~~,
\end{displaymath}
thus obtaining
\begin{displaymath}
I = Tr \int \prod_a (\frac{N}{2\pi} d\widehat{s}^a d\widehat{u}^a) 
\prod_{ab} (\frac{N}{2\pi} d\widehat{s}^{ab} d\widehat{t}^{ab}) 
\end{displaymath}
\begin{displaymath}
\times \exp~ \{ -N ~[ ~\frac{1}{2} \sum_a \vert \widehat{s}^a \vert ^2 +
 \frac{1}{2} \sum_a \vert \widehat{u}^a \vert ^2 
\end{displaymath}
\begin{displaymath}
+ \frac{1}{2} \sum_{a\neq b} (\vert \widehat{s}^{ab} \vert ^2 + \vert \widehat{t}^{ab} \vert ^2) 
- {\beta}^{\frac{1}{2}} \sum_a \widehat{s}^a \overrightarrow{S}^a -
\frac{\beta}{2} \widehat{u}^a \overrightarrow{U}^a  
\end{displaymath}
\begin{displaymath}
- \frac{\beta}{2} \sum_{a\neq b} ( \widehat{s}^{ab} \overrightarrow{S}^{ab} +
\widehat{t}^{ab} \overrightarrow{T}^{ab} )~]~ \}~~.
\end{displaymath}
The mean field approximation consists on saying that one can deal with independent spins 
feeling the external conjugated fields $\widehat{\blacksquare}$ so that this last relation becomes
\begin{displaymath}
I = \int \prod_a (\frac{N}{2\pi} d\widehat{s}^a d\widehat{u}^a) 
\prod_{ab} (\frac{N}{2\pi} d\widehat{s}^{ab} d\widehat{t}^{ab}) 
\end{displaymath}
\begin{displaymath}
\times ~Tr_n ~[~ \exp~ \{-\frac{1}{2} \sum_a \vert \widehat{s}^a \vert ^2 -
 \frac{1}{2} \sum_a \vert \widehat{u}^a \vert ^2 
\end{displaymath}
\begin{displaymath}
- \frac{1}{2} \sum_{a\neq b} (\vert \widehat{s}^{ab} \vert ^2 + \vert \widehat{t}^{ab} \vert ^2) 
+ {\beta}^{\frac{1}{2}} \sum_a \widehat{s}^a \overrightarrow{S}^a +
\frac{\beta}{2} \widehat{u}^a \overrightarrow{U}^a  
\end{displaymath}
\begin{equation}
+ \frac{\beta}{2} \sum_{a\neq b} ( \widehat{s}^{ab} \overrightarrow{S}^{ab} +
\widehat{t}^{ab} \overrightarrow{T}^{ab} )~ \}~~]^N
\label{free_en_3}
\end{equation}
where from now on the trace is over the $n$ replicas at a single site (mean field approximation).

\section{Spin-glass self-consistent equation and its numerical solution}

At this point we have to impose the following spin-glass constraints \cite{SK2}, 
based on the physical meaning of the two order parameters in the {\it M=0 QSS} regime (that we are considering as an effective spin-glass phase of the HMF model):
\begin{enumerate}
\item{
the first one refers to the $magnetization$, that is null in the {\it M=0 QSS} regime, thus implying
\begin{equation}
\vert \widehat{s}^a \vert \vert \widehat{u}^a \vert = 0
\end{equation}
}
\item{
the second one refers to the {\it spin-glass order parameter}, which of course does not vanish in the spin-glass phase thus quantifying the degree of freezing of the rotators in the {\it M=0 QSS} regime; as usual in standard replica method for glassy systems, it is chosen 
as proportional to the module of the $overlap$ between two different replicas at a single site
\begin{equation}
q_n = q^{ab} = 2 \beta^{-1} \vert \widehat{s}^{ab} \vert~~~,
\label{pn}
\end{equation}
where the $\vert \widehat{s}^{ab} \vert$ are considered equals $\forall {a,b}$ (Replica-Symmetry approximation). 
The phase of $\widehat{s}^{ab}$, being arbitrary, can be set to zero for convenience so that its direction is $\overrightarrow{u_x}$.
Finally, we set also $\vert \overrightarrow{t}^{ab} \vert = 0$.
}
\end{enumerate}
By incorporating the spin-glass constraints with delta functions 
one can easily perform the integral in Eq.(\ref{free_en_3}), thus obtaining
\begin{equation}
I = [~ Tr_n ~ \exp(~-\sum_{a\neq b} \frac{\beta {q_n}^2}{4} 
-\sum_{a\neq b} \frac{{\beta}^2 q_n}{4} \overrightarrow{u}_x \overrightarrow{S}^{ab} ~)~]^N~~~.
\end{equation}
Since the first term is independent on the replicas and the trace being a linear application, 
one has equivalently
\begin{displaymath}
I = exp(-N \sum_{a\neq b} \frac{\beta {q_n}^2}{4})[Tr_n~exp( -\sum_{a\neq b} 
\frac{{\beta}^2 q_n}{4} \overrightarrow{u}_x \overrightarrow{S}^{ab} ~)~]^N
\end{displaymath}
\begin{displaymath}
= exp(~-N [\frac{n(n-1)}{2} \frac{\beta {q_n}^2}{4} 
\end{displaymath}
\begin{equation}
- \ln Tr_n ~exp(-
\frac{{\beta}^2 q_n}{4} \sum_{a\neq b}\overrightarrow{u}_x \overrightarrow{S}^{ab} ~)~]~)~~.
\label{steepest}
\end{equation} 
Reminding  that $\overrightarrow{S}^{ab}$ is the replica interference at single site, i.e.
$\overrightarrow{S}^{ab}=(cos({\theta}^a - {\theta}^b),sin({\theta}^a - {\theta}^b))$, one has
that $\overrightarrow{u}_x \overrightarrow{S}^{ab}=cos({\theta}^a - {\theta}^b)\overrightarrow{s}^{a} \overrightarrow{s}^{b}$, so we can write
\begin{equation}
 \sum_{a\neq b}\overrightarrow{u}_x \overrightarrow{S}^{ab}=(\sum_a \overrightarrow{s}^{a})^2 - n~~.
\end{equation}
Inserting the latter relation in Eq.(\ref{steepest}), using the Hubbard-Stratonovich
transformation (for $exp[-\frac{{\beta}^2 q_n}{4} (\sum_a \overrightarrow{s}^{a})^2~$]) once again
and substituting the trace with the following multiple integral
\begin{displaymath}
Tr [...] = \int_{0}^{2\pi} \prod_{a=1}^n \frac{d\theta^a}{2\pi} [...]~~,
\end{displaymath}
one finally finds
\begin{displaymath}
I = \exp~ (-N ~[ \frac{n(n-1)}{2} {(\frac{\beta q_n}{2})}^2
\end{displaymath}
\begin{equation}
- ln~\int_{0}^{\infty} r~ dr \exp[- \frac{{r}^2}{2}
- \frac{n q_n \beta^2}{4}]~ {I_0}^n (\beta r \sqrt{q_n/2})~]~) ~~.
\label{exp_sd}
\end{equation}
The latter exponential, for $N\rightarrow \infty$, will show a maximum at the extremum of its argument,
thus we have to impose the following (steepest descent) extremal condition
\begin{displaymath}
\frac{\partial}{\partial q_n}~[ \frac{n(n-1)}{2} {(\frac{\beta q_n}{2})}^2
\end{displaymath}
\begin{displaymath}
- ln~\int_{0}^{\infty} r~ dr \exp[- \frac{{r}^2}{2}
- \frac{n q_n \beta^2}{4}]~ {I_0}^n (\beta r \sqrt{q_n/2})~] = 0~~.
\end{displaymath}
Performing the derivative we find the desired self-consistent 
equation for the order parameter $q_n$
\begin{displaymath}
q_n (1-n) = 1 - \sqrt{\frac{2}{q_n}}\beta^{-1} 
\end{displaymath}
\begin{displaymath}
\times \frac{\int_{0}^{\infty} r^2~ dr
\exp[- \frac{{r}^2}{2}]
~ {I_0}^n (\beta r \sqrt{q_n/2})~ \frac{I_1(\beta r \sqrt{q_n/2})}{I_0(\beta r \sqrt{q_n/2})}}
{\int_{0}^{\infty} r~ dr \exp[- \frac{{r}^2}{2}]~ {I_0}^n (\beta r \sqrt{q_n/2})}
\end{displaymath}
that has to be continued for $n\rightarrow0$, giving finally
\begin{equation}
q = 1 - \sqrt{\frac{2}{q}}\beta^{-1} ~ \int_{0}^{\infty} r^2~ dr
\exp[- \frac{{r}^2}{2}]
~ \frac{I_1(\beta r \sqrt{q/2})}{I_0(\beta r \sqrt{q/2})}
\label{p_equat}
\end{equation}

Solving numerically this equation, we can immediately calculate the expected value of the spin-glass order parameter $q$ as a function of the temperature $T=\beta^{-1}$ and compare it with the polarization order parameter $p$ of the HMF model. In particular, since we have to consider {\it M=0 QSS} with null magnetization, we  consider the limiting temperature $T_{QSS}$ (i.e. the temperature for $N\longrightarrow \infty$) and, by varying it in Eq.(\ref{p_equat}), we  obtain the theoretical curve  shown in Fig.\ref{fig2} (full line). This curve is compared with the  molecular dynamics simulations for the polarization (open circles) performed in the QSS regime of the HMF model for N=1000 and for different temperatures in the subcritical region. Some corresponding energy density values are also reported for convenience. As previously stressed, these results are independent on N
within the error bars \cite{kyoto} and this allow us to extrapolate them to the thermodynamic limit.
\\
The theoretical curve predicts, as expected,  a phase transition at $T_c=1/2$ and superimposes on the values of the polarization obtained in the range $0.36<T_{QSS}<0.45$, that corresponds (being there $M_{QSS}=0$) to the energy density range $0.68<U<0.72$ through the zero magnetization (minimum temperature) line equation  $U=\frac{T}{2}+\frac{1}{2}$ , see Eq.(\ref{caloric}).
The simulations points start to disagree with the results of the theoretical curve around $T_{QSS}\sim 0.45$, i.e. for $U > 0.72$: above these values the QSS points are very close to the correspondent equilibrium temperature values on the caloric curve (as visible in the inset of Fig.\ref{fig2}) and the glassy features of the QSS regime tend to disappear. 
Inverting the previous argument, we could also suggest that such a theoretical result allows us to better specify the range of energy densities where the QSS regime can be considered as a spin-glass phase for the HMF model.

\section{Conclusions} 

We have discussed the glassy phase of the HMF model by means of an effective spin-glass Hamiltonian and shown that the corresponding order parameter coincides with the polarization  in the energy range where HMF exhibits strong dynamical anomalies and glassy dynamics.
Being a long-range Hamiltonian solvable model and showing the existence of a glassy-like relaxation dynamics along quasi-stationary trajectories, the HMF model poses  new challenging questions on the origin of its glassy unexpected behavior.
In these respects, the model may be a very useful "laboratory" for studying  general trends that can be later tested numerically and experimentally 
in more realistic systems and real materials.
\\
These analytical results, obtained within the Replica-Symmetry SK framework and in the thermodynamic limit, seem to give further support to the interpretation of the QSS regime as a real spin-glass phase. This is true at least in the limiting temperature region where the glassy features are more evident. In this region the elementary polarization 
seems to play the role of the EA order parameter in giving a measure of the numerically observed quenched dynamics. 
\\
In conclusion, we do hope that this connection   between the HMF model and Spin-glass systems could bring new insight on the several common features.

\section{Acknowledgments}

We  thank Andrea De Martino, Andrea Giansanti and Irene Giardina for many useful and 
stimulating discussions.



\begin{references}
\bibitem{numerics} 
A list of references where 
the nonequilibrium evolution of different glassy systems has been studied 
numerically can be found in  ref.\cite{Cugliand}.
\end{references}
\end{document}